\documentclass{ws-ijmpcs}

\begin{document}

\markboth{Cecilia Chirenti, Patrick R. Silveira and Odylio D. Aguiar}
{Non-radial oscillations of neutron stars and the detection of gravitational waves}

%%%%%%%%%%%%%%%%%%%%% Publisher's Area please ignore %%%%%%%%%%%%%%%
%
%\catchline{}{}{}{}{}
%
%%%%%%%%%%%%%%%%%%%%%%%%%%%%%%%%%%%%%%%%%%%%%%%%%%%%%%%%%%%%%%%%%%%%

\title{NON-RADIAL OSCILLATIONS OF NEUTRON STARS AND THE DETECTION OF GRAVITATIONAL WAVES}

\author{CECILIA CHIRENTI}

\address{Center for Mathematics, Computation and Cognition\\Federal University of ABC, Rua Santa Adélia, 166\\ Santo Andr\'e, 09210-170, Brazil\\
cecilia.chirenti@ufabc.edu.br}

\author{PATRICK R. SILVEIRA}

\author{ODYLIO D. AGUIAR}

\address{Astrophysics Division, National Institute for\\Space Research,
Av. dos Astronautas, 1758\\ S\~ao Jos\'e dos Campos, 12207-010, Brazil\\
patrick.silveira@das.inpe.br\\odylio.aguiar@das.inpe.br}

\maketitle

\begin{history}
\received{Day Month Year}
\revised{Day Month Year}
\end{history}

\begin{abstract}
We study the non-radial oscillations of relativistic neutron stars, in particular the (fundamental) f-modes, which are believed to be the most relevant for the gravitational wave emission of perturbed isolated stars. The expected frequencies of the f-modes are compared to the sensitivity range of Mario Schenberg, the Brazilian gravitational wave spherical detector.

\end{abstract}

\ccode{PACS numbers: 04.40.Dg, 95.30.Sf}

\section{Introduction}	
Why study oscillations of neutron stars? When it comes to sources of gravitational waves, binary systems of compact objects are among the most promising sources. But after a binary system inspiralls and merges, we are left with one perturbed compact object, which will somehow radiate away its extra energy to achieve a new equilibrium state, including gravitational radiation. And this final compact object could even be a super-massive neutron star, depending on the initial components of the binary system and the equation of state of the neutron star. 
Also single (isolated) neutron stars can be perturbed due to the interplay between the fluid of which the star is composed, its solid crust and magnetic field, for instance. Of course, this is a very complicated problem, beyond the scope of our present paper. But any generic perturbation with $\ell \ge 2$ can lead to the generation of gravitational waves. 
The stellar modes of oscillation are classified according to the restoring force acting on them \cite{KS}. They form different families of modes, with separated frequencies. For a non-rotating star (without magnetic fields) the usual classification gives:

\begin{itemize}
\item $f$-modes (fundamental): frequency is proportional to the square root of the mean density of the star. These are favoured for the gravitational wave emission.

\item $p$-modes (pressure): pressure is the restoring force, and the frequencies are higher than those of the f-modes.

\item $g$-modes (gravity): buoyancy is the restoring force, frequencies are lower than those of the f-modes.
\end{itemize}

For the spherical antennas, such as the Brazilian Mario Schenberg and the Dutch Mini-GRAIL \cite{Aguiar}, the $f$-modes would be particularly interesting due to the possibility of detection inside the 2.8-3.4 kHz frequency range, where they have their sensitivity bands.

This paper is organized as follows: in section \ref{sec:eq} we present our equilibrium stellar model and in section \ref{sec:pert} we review the perturbation equations for the $f$-modes. We discuss the numerical method and present our results in section \ref{sec:num} and conclude with our final remarks in section \ref{sec:final}.

\section{Equilibrium Star}
\label{sec:eq}

We begin with the spherically symmetric metric
\begin{equation}
dS^2 = -e^{\nu(r)}dt^2 + e^{\lambda(r)}dr^2 + r^2(d\theta^2 + \sin^2\theta d\phi^2),
\end{equation}
and the perfect fluid stress-energy tensor
\begin{equation}
T^{\mu\nu} = (\epsilon + p)u^{\mu}u^{\nu} + pg^{\mu\nu},
\end{equation}
where $\epsilon$ is the total energy density, $p$ is the pressure and $u^{\mu}$ is the fluid 4-velocity to write the TOV equations
\begin{equation}
m' = 4\pi \epsilon r^2\,,\quad 
\nu' = \frac{8\pi p r^3 + 2m}{r\left(r-2m\right)}\,,\quad
p' = -\nu'\frac{(\epsilon + p)}{2}\,.
\end{equation}

Finally, we use a polytropic equation of state $p = K\rho^{\Gamma}$, where $\rho$ is the rest-mass energ density of the fluid \cite{Tooper}.

\section{Polar Perturbations}
\label{sec:pert}

The linearized Einstein equations
\begin{equation}
\delta\left(R_{\mu \nu} - \frac{1}{2}g_ {\mu \nu}R\right) = 8\pi\delta
T_{\mu \nu}\quad \textrm{and} \quad \delta(T^{\mu}_{\phantom{\mu}\nu;\mu}) = 0\,,
\end{equation}
with the following Ansatz (polar perturbations) for the perturbed metric tensor 
\begin{eqnarray}
dS^2 &=& -e^{\nu}(1+r^{\ell}H_0Y^{\ell}_{m}e^{i\omega t})dt^2 
- 2i\omega r^{\ell+1}H_1Y^{\ell}_me^{i\omega t}dtdr + \nonumber\\
&+& e^{\lambda}(1 - r^{\ell}H_0Y^{\ell}_{m}e^{i \omega t})dr^2 
+ r^2(1 - r^{\ell}KY^{\ell}_{m}e^{i \omega t})(d\theta^2 + \sin^2\theta d\phi^2)\,,
\end{eqnarray}
and for the perturbation of the fluid in the star \cite{TC1967} (Lagrangian displacement)
\begin{equation}
\xi_r = r^{\ell-1}e^{\lambda/2}WY^{\ell}_{m}e^{i\omega t}\,,\quad
\xi_{\theta} = -r^{\ell}V\partial_{\theta}Y^{\ell}_{m}e^{i\omega t}\,,\quad
\xi_{\phi} = -r^{\ell}V\partial_{\phi}Y^{\ell}_{m}e^{i\omega t}
\end{equation}
lead to a 4th order system:
\begin{eqnarray}
H_1' &=&
 -\frac{1}{r} \biggl[ \ell+1+\frac{2Me^\lambda}{r}+4\pi
  r^2e^\lambda(p-\epsilon) \biggr]
 + \frac{e^\lambda}{r}
 \left[ H_0 + K - 16\pi(\epsilon+p)V \right] \:, \nonumber \\
 K' &=&
 \frac{1}{r} H_0 + \frac{\ell(\ell+1)}{2r} H_1
 - \left[ \frac{\ell+1}{r}+\frac{\nu'}{2} \right] K
 - 8\pi(\epsilon+p)\frac{e^{\lambda/2}}{r} W \:, \nonumber \\
 W' &=&
 - \frac{\ell+1}{r} W 
 + re^{\lambda/2} \left[ \frac{e^{-\nu/2}}{(\epsilon+p)c_s^2} X
 - \frac{\ell(\ell+1)}{r^2} V + \frac{1}{2}H_0 + K \right] \:, \nonumber \\
 X' =
 &-&\frac{\ell}{r} X + \frac{(\epsilon+p)e^{\nu/2}}{2}
 \Biggl[ \left( \frac{1}{r}+\frac{\nu'}{2} \right)
 + \left(r\omega^2e^{-\nu} + \frac{\ell(\ell+1)}{2r}\right) H_1
 + \left(\frac{3}{2}\nu' - \frac{1}{r}\right) K \nonumber \\
&-& \frac{\ell(\ell+1)}{r^2}\nu' V 
 - \frac{2}{r} 
 \Biggl( 4\pi(\epsilon+p)e^{\lambda/2} + \omega^2e^{\lambda/2-\nu}
 - \frac{r^2}{2}
 \biggl(\frac{e^{-\lambda/2}}{r^2}\nu'\biggr)' \Biggr) W \Biggr] \:.
\end{eqnarray}

where $X = \omega^2(\epsilon+p)e^{-\nu/2}V - \frac{p'}{r}e^{(\nu-\lambda)/2}W + \frac{1}{2}(\epsilon+p)e^{\nu/2}H_0$ and $H_0$ is given by another algebraic relation \cite{LD1983,LD1985}.

\section{Numerical Method and results for the $f$-modes}
\label{sec:num}

We look for solutions that are regular at the origin $r=0$,
have vanishing perturbed pressure at the surface $r = R$
and describe only outgoing gravitational waves at infinity $r \to \infty$.

Our numerical code follows closely the prescriptions of Ref. \refcite{LD1985}. We use as an initial guess the Newtonian frequency of the f-modes: 
\begin{equation*}
\omega = \sqrt{\frac{2\ell(\ell-1)}{2\ell+1}\left(\frac{M}{R^3}\right)}\,,
\end{equation*}
and we solve the eigenvalue problem with a modified shooting method. By combining at $R/2$ the two linearly independent regular solutions at $r=0$ with the three linearly independent solutions with vanishing pressure ($X(R) = 0$) at the surface we create the initial conditions for solving the Zerilli equation outside of the star, imposing then only outgoing waves at infinity.

\begin{figure}[pb]
\centerline{\psfig{file=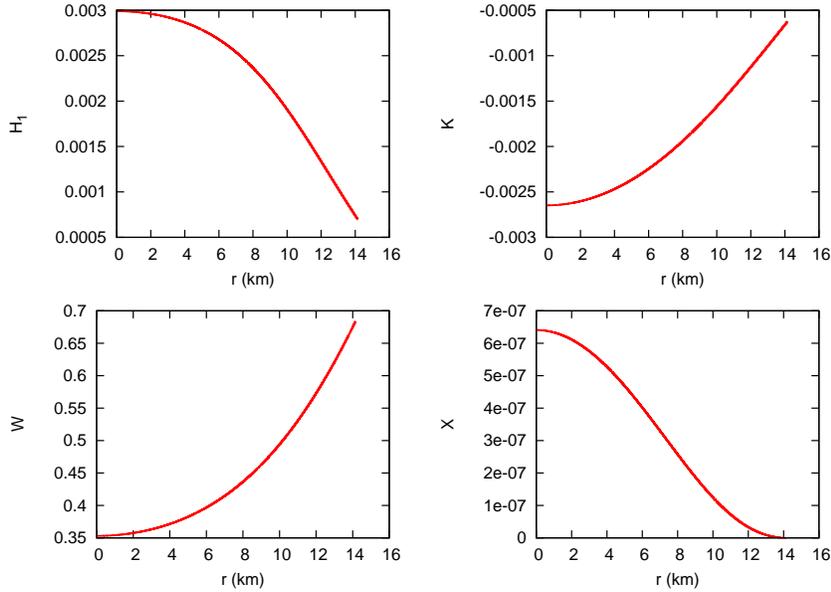,width=8.0cm,angle=-90}}
\vspace*{8pt}
\caption{Behavior of the perturbation functions inside the star. 
\label{f2}}
\end{figure}

\begin{figure}[pb]
\centerline{\psfig{file=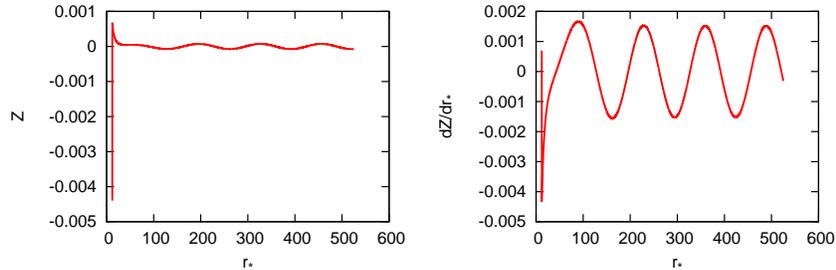,width=4.0cm,angle=-90}}
\vspace*{8pt}
\caption{Perturbations outside the star, given in terms of the Zerilli function $Z$ and its derivative, as functions of the tortoise coordinate $r_*$. 
\label{f3}}
\end{figure}

In Figs. \ref{f2} and \ref{f3} we show the perturbation functions inside and outside the star for a typical case (with $K = 100$, $\Gamma = 2$ and $\rho = 1.28 \times 10^{-3}$ in the equation of state, given here in code units, i.e. $c=G=M_{\textrm{sun}}=1$).

Typical values for the complex frequency $\omega = 2\pi f + i/\tau$ of the $f$-modes for a neutron star have $f = 1.5-3$ kHz and $\tau = 0.1-0.5$ s \cite{KS}. The values obtained for typical star shown in Figs. \ref{f2} and \ref{f3} are $f = 1.5795$ kHz and $\tau = 0.2987$ s.

\begin{figure}[pb]
\centerline{
\psfig{file=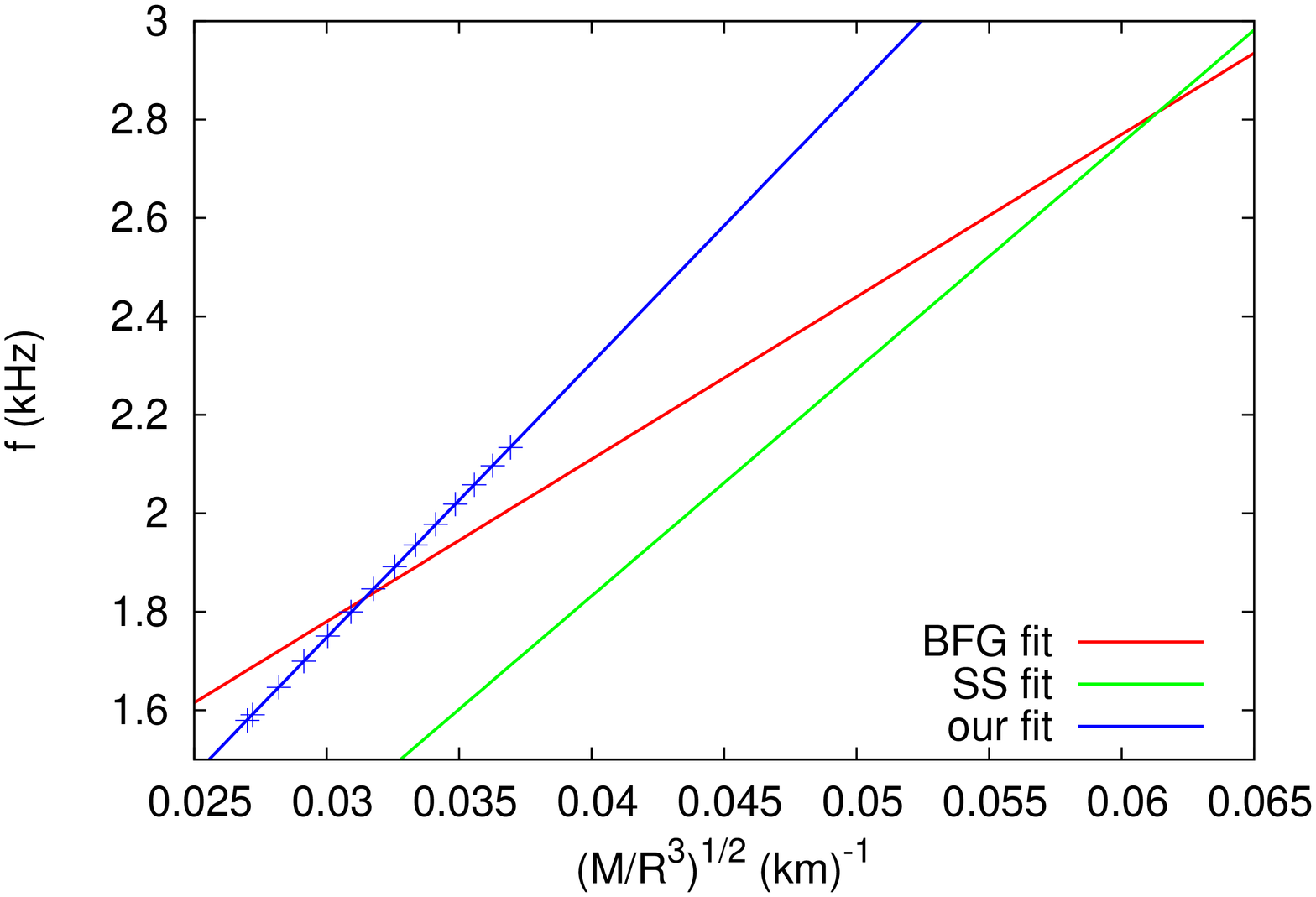,width=4.0cm,angle=-90}
\psfig{file=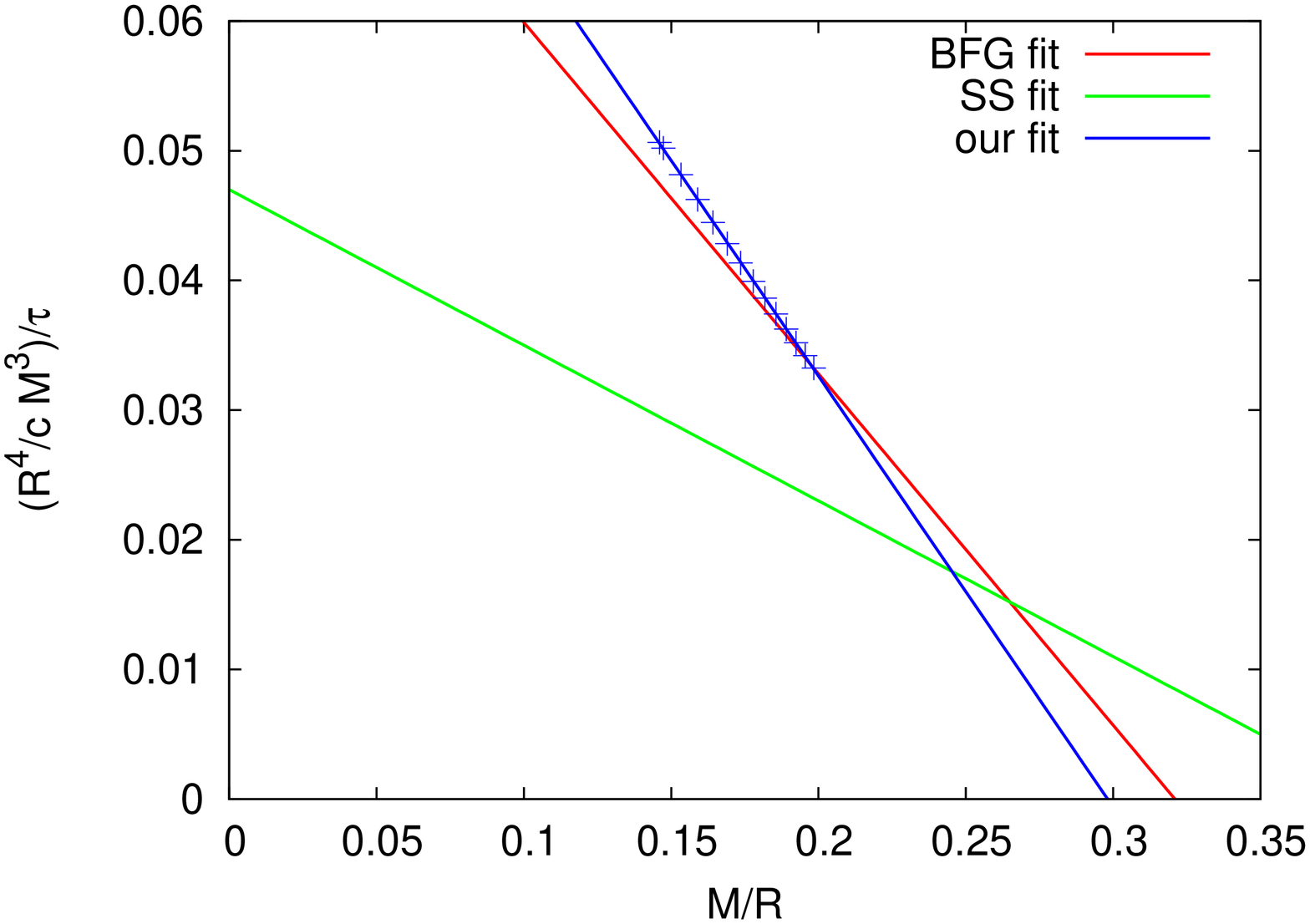,width=4.0cm,angle=-90}}
\vspace*{8pt}
\caption{Numerical fits for the frequency and damping time of the $f$-modes.
\label{f4}}
\end{figure}

It has been observed that $f$ and $\tau$ can be fitted by simple expressions in terms of the total mass $M$ and the radius $R$ of the star \cite{AK,BFG}. For a sequence of stars with our typical polytropic equation of state and increasing central density, we produced the following fit:
\begin{equation}
f = 7.36 \times 10^{-2} + 55.80 \sqrt{\frac{M}{R^3}}\quad \textrm{and}\quad
\frac{1}{\tau} = \frac{M^3}{R^4}\left[9.91\times10^{-2} -0.33\left( \frac{M}{R}\right)\right]\,,
\end{equation}
which can be seen in Fig. \ref{f4}, together with two other fits from Ref. \refcite{BFG}, that used realistic equations of state and equations of state containing strange quarks \cite{VF}.

\section{Observational Consequences and Final Remarks}
\label{sec:final}

Ideally, after observing the $f$-mode frequencies from neutron star emitting gravitational waves, we will be able to solve the initial value problem and determining the total mass and radius of the star. With this information, we will gain precious information on the neutron star equation of state, which is still largely unknown, and object of investigations from different branches of physics. 

From the observational point of view, the $f$-modes are the most relevant ones, expected to be favoured for the gravitational wave emission. Our fits for the frequencies, even though obtained used only a very simple polytropic equation of state, already present the expected behavior and typical values. 
The Mario Schenberg and Mini-GRAIL spherical antennas may observe these neutron star $f$-modes in a near future, if some of these modes fall inside the 2.8-3.4 kHz frequency range, where they have their sensitivity bands.

\section*{Acknowledgments}

The authors are thankful to Lee Lindblom, Leonardo Gualtieri, Luciano Rezzolla and Shinichiro Yoshida for useful discussions, help and encouragement in different stages of this project. This work was supported by the Brazilian agencies FAPESP, CNPq and CAPES, and by the Max Planck Society.

\end{document}